\documentclass[11pt,draftcls,onecolumn]{IEEEtran}

\usepackage{amsmath,url}
\usepackage{epsfig,amssymb,amsbsy,verbatim,array}
\usepackage{pstricks,psfrag,theorem,cite}

\let\intern=\iftrue

\def\argmax{\operatorname{arg~max}}
\def\figref#1{Fig.\,\ref{#1}}%
\def\E{\mathbb{E}}
\def\P{\mathbb{P}}
\def\R{\mathbb{R}}
\def\W{\mathcal{W}}

\def\L{\mathcal{L}}
\def\Psi{\mathcal{L}}
\def\ie{{\em i.e.}}
\def\eg{{\em e.g.}}
\def\V{\operatorname{Var}}

\def\erf{\operatorname{erf}}
\def\dilog{\operatorname{dilog}}
\def\Ei{\operatorname{Ei}}
\def\d{\mathrm{d}}
\def\opt{\mathrm{opt}}

\def\SIR{\mathrm{SIR}}
\def\one{\mathbf{1}}

\newtheorem{proposition}{Proposition}



\begin{document}
\title{Outage and Local Throughput and Capacity\\of Random Wireless
Networks}
\author{Martin Haenggi,~\IEEEmembership{Senior Member,~IEEE}}
\maketitle

\begin{abstract}
Outage probabilities and single-hop throughput are two important performance
metrics that have been evaluated for certain specific types of wireless networks. 
However, there is a lack of comprehensive results for larger classes of networks,
and there is no systematic approach that permits the convenient comparison of the
performance of networks with different geometries and levels of randomness.

The {\em uncertainty cube} is introduced to categorize the uncertainty
present in a network. The three axes of the cube represent the three main
potential sources of uncertainty in interference-limited networks: the node
distribution, the channel gains (fading), and the channel access (set of transmitting nodes). 
For the performance analysis, a new parameter, the so-called
{\em spatial contention}, is defined. It measures the slope of the outage
probability in an ALOHA network as a function of the transmit probability $p$
at $p=0$. Outage is defined as the event that the signal-to-interference ratio (SIR)
is below a certain threshold in a given time slot.
It is shown that the spatial contention is sufficient to characterize
outage and throughput in large classes of wireless networks, corresponding to
different positions on the uncertainty cube. Existing
results are placed in this framework, and new ones are derived.

Further, interpreting the outage probability as the SIR distribution,
the ergodic capacity of unit-distance links is determined and compared
to the throughput achievable for fixed (yet optimized) transmission rates.
\end{abstract}

\section{Introduction}

\subsection{Background}
In many large wireless networks, the achievable performance is limited
by the interference. Since the seminal paper \cite{net:Gupta00} the {\em scaling behavior} of
the network throughput or transport capacity has been the subject of intense
investigations, see, \eg,  \cite{net:Xue06now}
and references therein. Such ``order-of" results are certainly important but do not
provide design insight when different protocols lead to the same scaling behavior.
On the other hand, relatively few {\em
quantitative} results on outage and local (per-link) throughput are available.
While such results provide only a microscopic view of the network, we can
expect concrete performance measures that permit, for example, the fine-tuning
of channel access probabilities or transmission rates.
Using a new parameter termed {\em spatial contention},
we classify and extend the results in
\cite{net:Sousa90,net:Linnartz92,net:Mathar95wn,net:Baccelli06} to
general stochastic wireless networks with up to three dimensions of uncertainty:
node placement, channel characteristics, and channel access.

\subsection{The uncertainty cube}
The level of uncertainty of a network is determined by its
position in the {\em uncertainty cube}.
The three coordinates $(u_l,u_f,u_a)$, $0\leqslant u_l,u_f,u_a\leqslant 1$, denote the
degree of uncertainty in the node placement, the channels, and the
channel access scheme, respectively. Values of 1
indicate complete uncertainty (and independence), as specified in
Table \ref{table:uncertainty}.
\begin{table}[h]
\normalsize
\centering
\begin{tabular}{|l|l|l|}
\hline
Node location & $u_l=0$ & Deterministic node placement\\
 & $u_l=1$ & Poisson point process\\\hline
Channel (fading) & $u_f=0$ & No fading\\
     & $u_f=1$ & Rayleigh (block) fading\\\hline
Channel access & $u_a=0$ & TDMA\\
   & $u_a=1$ & slotted ALOHA\\\hline
\end{tabular}
\caption{Specification of the uncertainty cube.}
\label{table:uncertainty}
\end{table}
The value of the $u_f$-coordinate corresponds to the fading
figure (amount of fading). For the Nakagami-$m$ fading model, for example,
we may define $u_f\triangleq 1/m$. A network with $(u_l,u_f,u_a)=(1,1,1)$
has its nodes distributed according to a Poisson point
process (PPP), all channels are Rayleigh (block) fading, and the channel
access scheme is slotted ALOHA.  The other extreme would be the $(0,0,0)$
network where the node's positions are deterministic, there is no fading, and
there is a deterministic scheduling mechanism.
Any point in the unit cube
corresponds to a meaningful practical network---the three axes are independent.
Our objective is
to characterize outage and throughput for the relevant corners of
this uncertainty cube. 

We focus on the interference-limited case, so we do not consider
noise\footnote{In the Rayleigh fading case, the outage expressions
factorize into a noise part and an interference part, see \eqref{laplace}.
So, the noise
term is simply a multiplicative factor to $p_s$.}. It is assumed that
all nodes transmit at the same power level that can be set to 1 since only
relative powers matter.  The performance results are
also independent of the absolute scale of the
network since only relative distances matter.

\subsection{Models, notation, and definitions}
{\em Channel model.} For the large-scale path loss (deterministic
channel component), we assume the standard power law where the received
power decays with $r^{-\alpha}$ for a path loss exponent
$\alpha$.  If all channels are Rayleigh, this is sometimes referred to
as a ``Rayleigh/Rayleigh'' model; we denote this case as ``1/1" fading. If either only the desired
transmitter or the interferers are subject to fading, we speak of {\em
partial fading}, denoted as ``1/0" or ``0/1" fading, respectively.

{\em Network model.} We consider a single link of distance
1, with a (desired) transmitter and receiver in a large network with
$n\in\{1,2,\ldots,\infty\}$ other nodes as potential interferers. 
The signal power (deterministic channel) or average signal power
(fading channel) at the receiver is 1. 
The distances to the interferers are denoted by $r_i$.
In the case of a PPP as the node distribution, the intensity is $1$.
For regular line networks, the inter-node distance is $1$.

{\em Transmit probability $p$.} In slotted ALOHA, every node
transmits independently with probability $p$ in each timeslot. Hence if the
nodes form a PPP of unit intensity,
the set of transmitting nodes in each timeslot
forms a PPP of intensity $p$.
The interference from node $i$ is $I_i=B_i G_i r_i^{-\alpha}$,
where $B_i$ is iid Bernoulli with parameter $p$ and $G_i=1$ (no fading) or
$G_i$ is iid exponential with mean 1 (Rayleigh fading).

{\em Success probability $p_s$.} A transmission is successful if the
channel is not in an outage, \ie, if the (instantaneous) SIR
$S/I$ exceeds a certain threshold $\theta$:
$p_s=\P[\SIR>\theta]$, where $I=\sum_{i=1}^n I_i$.  This is the
reception probability given that the desired transmit-receiver pair
transmits and listens, respectively.

{\em Effective distances $\xi_i$.} The {\em effective
distance $\xi_i$} of a node to the receiver is defined as
$\xi_i\triangleq r_i^\alpha/\theta$.

{\em Spatial contention $\gamma$} and {\em spatial efficiency $\sigma$.} 
For a network using ALOHA with transmit probability $p$, define
\begin{equation}
 \gamma\triangleq-\frac{\d p_s(p)}{\d p}\Big|_{p=0}\,,
\end{equation}
\ie, the slope of the outage probability $1-p_s$ at $p=0$,
as the {\em spatial contention} measuring how concurrent transmissions (interference)
affect the success probability.
$\gamma$ depends on the SIR threshold $\theta$,
the geometry of the network, and the path loss exponent $\alpha$. Its inverse
$\sigma\triangleq 1/\gamma$ is the
{\em spatial efficiency} which quantifies how efficiently a network uses space as a resource.

{\em (Local) probabilistic throughput $p_T$.} The probabilistic throughput
is defined to be the success probability
multiplied by the probability that the transmitter actually transmits (in full-duplex
operation) and, in addition in half-duplex operation, the receiver actually listens.
So it is the unconditioned reception probability. This
is the throughput achievable with a simple ARQ scheme (with error-free
feedback) \cite{net:Ahmed03allerton}.
For the ALOHA scheme, the half-duplex probabilistic throughput is
$p_T^h\triangleq p(1-p)p_s$ and for full-duplex it is $p_T^f=p\,p_s$.
For a TDMA line
network where nodes transmit in every $m$-th timeslot,
$p_T\triangleq p_s/m$.

{\em Throughput $T$.} The throughput is defined as the product of the
probabilistic throughput and the rate of transmission, assuming that capacity-achieving
codes are used, \ie, $T\triangleq p_T \log(1+\theta)$.

{\em Ergodic capacity $C$.} Finally, interpreting $1-p_s(\theta)$ as the distribution of the
$\SIR$, we calculate $C\triangleq \E\log(1+\SIR)$.

\section{Related Work}

The study of outage and throughput performance is related to the problem of
interference characterization. Important results on the interference in large
wireless systems have been derived by \cite{net:Lowen90,net:Sousa90b,net:Mathar95wn,net:Hellebrandt97wn,net:Ilow98}.
In \cite{net:Linnartz92}, outage probabilities for cellular networks are calculated for
channels with Rayleigh fading and shadowing while \cite{net:Sousa90} determines
outage probabilities to determine the optimum transmission range in a Poisson network.
\cite{net:Zorzi95} combined the two approaches to determine the optimum transmission
range under Rayleigh fading and shadowing.
\cite{net:Baccelli06} provides a detailed analysis on outage probabilities and routing progress
in Poisson networks with ALOHA.

For our study of $(1,0,1)$, $(0,1,1)$, and $(1,1,1)$ networks, we will draw on results from
\cite{net:Sousa90,net:Mathar95wn,net:Zorzi95,net:Baccelli06}, as discussed in the rest of this section.

\subsection{$(1,0,1)$: Infinite non-fading random networks with $\alpha=4$ and slotted ALOHA}
This case is studied in \cite{net:Sousa90}. The characteristic
function of the interference is determined to be\footnote{Note that
their notation is adapted to ours. Also, a small mistake in
\cite[Eqn. (18)]{net:Sousa90} is corrected here.}
\begin{align}
 \E e^{j\omega I}&=\exp\bigl(-\pi p\Gamma(1-2/\alpha)e^{-j\pi/\alpha}\omega^{2/\alpha}\bigr)
 \label{sousa_gen}\\
\intertext{and, for $\alpha=4$,}
 &=\exp\bigl(-\pi\sqrt{\pi/2}(1-j) p\sqrt{\omega}\bigr)\,.
  \label{sousa_laplace}
\end{align}

\subsection{$(0,1,1)$: Regular fading networks with $\alpha=2$ and
slotted ALOHA}
In \cite{net:Mathar95wn}, the authors derive the distribution of the
interference power for one- and two-dimensional Rayleigh fading
networks with slotted ALOHA and $\alpha=2$.  Closed-form expressions
are derived for infinite regular line networks with $r_i=i$, $i\in\mathbb{N}$. 
The Laplace transform of the interference
is \cite[Eqn. (8)]{net:Mathar95wn}
\begin{equation}
 \L_I(s)=\frac{\sinh\bigl(\pi\sqrt{s(1-p)}\bigr)}{\sqrt{1-p}
   \sinh\bigl(\pi\sqrt{s}\bigr)}\,. 
\label{mathar}
\end{equation}

The Laplace transforms of the interference are particularly convenient for the
determination of outage probabilities in Rayleigh fading.
As was noted in \cite{net:Linnartz92,net:Zorzi95,net:Baccelli06}, the success
probability $p_s$ can be expressed as the product of the Laplace transforms
of the interference and noise:
\begin{align}
 p_s=\int_0^\infty e^{-s \theta} \d\P[N+I\leqslant s]
 =\Psi_I (\theta)\cdot \Psi_N (\theta)\,.
\label{laplace}
\end{align}

In the interference-limited regime, the Laplace transform of the interference itself
is sufficient. Otherwise an exponential factor for the noise term (assuming noise
with fixed variance) needs to be added.

\subsection{$(1,1,1)$: Random fading networks with slotted ALOHA}
In \cite{net:Zorzi95,net:Baccelli06}, \eqref{laplace} was calculated for a two-dimensional
random network with Rayleigh fading and ALOHA. Ignoring the noise,
they obtained (see \cite[Eqn. (3.4)]{net:Baccelli06},\cite[(Eqn. (A.11)]{net:Zorzi95})
\begin{equation}
 p_s=e^{-p \theta^{2/\alpha}C_2(\alpha)} 
\label{baccelli}
\end{equation}
with 
\begin{equation}
   C_2(\alpha)=\frac{2\pi\Gamma(2/\alpha)\Gamma(1-2/\alpha)}{\alpha}=
      \frac{2\pi^2}{\alpha}\csc\left(\frac{2\pi}{\alpha}\right)\,.
\end{equation}
The subscript 2 in $C_2$ indicates that this is a constant for the two-dimensional case.
Useful values include $C_2(3)=4\pi^2/3\sqrt{3}\approx 7.6$ and
$C_2(4)=\pi^2/2\approx 4.9$. $C_2(2)=\infty$,
so $p_s\rightarrow 0$ as $\alpha\rightarrow 2$ for any $\theta$. 
The spatial contention is $\gamma=\theta^{2/\alpha}C_2(\alpha)$.

\section{The Case of a Single Interferer}
To start, we consider the case of a single interferer at  effective
distance $\xi=r^\alpha/\theta$ transmitting with probability $p$, which is
the simplest case of a $(0,u_f,1)$-network. For the fading, we allow the
desired channel and the interferer's channel to be fading or static. If
both are Rayleigh fading (this is called the $1/1$ case),
the success probability is
\begin{equation}
p_s^{1/1}=\P[\SIR>\theta]=1-\frac{p}{1+\xi}\,. 
\label{ps11}
\end{equation}
For a fading desired link and non-fading interferers (denoted as $1/0$ fading),
$I=B r^{-\alpha}$ with $B$ Bernoulli with parameter $p$ and thus
\begin{align}
p_s^{1/0}&=\P[S>B/\xi]
 =1-p(1-e^{-1/\xi})\,.
 \label{ps10}
\end{align}

In the case of $0/1$ fading (non-fading desired link, fading interferer),
\begin{equation}
 p_s^{0/1}=\P[I < \theta^{-1}]=1-pe^{-\xi}\,.
 \label{ps01}
\end{equation}

For comparison, transmission success in the non-fading  ($0/0$) case is
guaranteed if $\xi>1$ or the interferer does not transmit, \ie, $p_s^{0/0}=1-p\one_{\xi\leqslant 1}$.

Hence in all cases the outage probability $1-p_s(p)$ is increasing linearly in $p$ with
slope $\gamma$. The values of $\gamma$ are summarized in Table \ref{table:single}.
\begin{table}
\begin{center}
\begin{tabular}{|l|l|}
\hline
Case & Spatial contention $\gamma$ \\\hline
1/1 & $\frac{1}{1+\xi}$\\
1/0 & $1-\exp(-1/\xi)$\\
0/1 & $\exp(-\xi)$\\
0/0 & $\one_{\xi\leqslant 1}$ \\
\hline
\end{tabular}
\end{center}
\caption{Spatial contention $\gamma$ in the single-interferer case.}
\label{table:single}
\end{table}

The ordering is $\gamma^{1/0}\geqslant \gamma^{1/1}\geqslant \gamma^{0/1}$,
with equality only if $\xi=0$, corresponding to an interferer at distance $0$ that causes
an outage whenever it transmits, in which case all $\gamma$'s  are one. 
The statement that $1-\exp(-1/\xi) > (1+\xi)^{-1}$, $\xi>0$ is the same as
$\log(1+\xi)-\log \xi < 1/\xi$, which is evident from interpreting the left side
as the integral of $1/x$ from $\xi$ to $1+\xi$ and the right side its Riemann
upper approximation $1/x$ times $1$. The ordering can also be shown using
Jensen's inequality: $\gamma^{1/0}\geqslant \gamma^{1/1}$ since 
$\E(\exp(-I\theta))\geqslant\exp(-\theta\E I)$ due to the convexity of the exponential.
And $\gamma^{1/1}\geqslant\gamma^{0/1}$ since
$\E(1-\exp(-S\xi))<1-\exp(-\xi\E S)$ due to the concavity of $1-\exp x$.
To summarize:

\begin{proposition}
In the single-interferer case, fading in the desired link is harmful whereas fading in the
channel from the interferer is helpful.
\end{proposition}

We also observe that for small $\xi$,
$\gamma^{1,1}\lessapprox \gamma^{0,1}$, whereas for larger $\xi$,
$\gamma^{1,1}\gtrapprox \gamma^{1,0}$. So if the interferer is relatively
close, it does not matter whether the desired link is fading or not.
On the other hand, if the interferer is relatively large, it hardly matters whether
the interferer's channel is fading.

The results can be generalized to Nakagami-$m$ fading in a straightforward manner.
If the interferer's channel is Nakagami-$m$ fading, while the desired link
is Rayleigh fading, we obtain
\begin{equation}
 p_s^{1/m^{-1}}=1-p\left(1-\frac{m^m}{(\xi^{-1}+m)^m}\right)\,.
\end{equation}
As a function of $m$, this is decreasing for all $\theta>0$, and in the limit converges
to $p_s^{1/0}$ as $m\rightarrow\infty$ (see \eqref{ps10}).
On the other hand, if the desired link is Nakagami-$m$, the success probability is
\begin{equation}
 p_s^{m^{-1}/1}=1-p\left(\frac{m\xi^{-1}}{1+m\xi^{-1}}\right)^m\,
\end{equation}
which {\em increases} as $m$ increases for fixed $\theta>0$ and approaches
\eqref{ps01} as $m\rightarrow\infty$.

The three success probabilities $p_s(\theta)$ are the complemetary cumulative
distributions (ccdf) of the SIR.

\section{Networks with Random Node Distribution}

\subsection{$(1,1,1)$: One-dimensional fading random networks with slotted ALOHA}
Evaluating \eqref{laplace} in the one-dimensional (and noise-free) case yields
\begin{equation}
 p_s=\exp\left(-\int_0^\infty \frac{2p}{1+r^\alpha/\theta}\d r\right)=\exp(-p\theta^{1/\alpha}C_1
 (\alpha))\,,
 \label{ps-oned}
\end{equation}
where $C_1(\alpha)=2\pi\csc(\pi/\alpha)/\alpha$. For finite $C_1$, $\alpha>1$ is needed.
$C_1(2)=\pi$, $C_1(4)=\pi/\sqrt{2}=\sqrt{C_2(4)}$. So the spatial contention is
$\gamma=\theta^{1/\alpha}C_1(\alpha)$.
For a general $d$-dimensional network, we may conjecture that
$\gamma=\theta^{d/\alpha}C_d(\alpha)$, with $C_d=c_d (d\pi/\alpha)\csc(d\pi/\alpha)$ and
$c_d\triangleq \pi^{d/2}/\Gamma(1+d/2)$
the volume of the $d$-dim.~unit ball. $\alpha>d$ is necessary for finite $\gamma$.
This generalization is consistent with \cite{net:Haenggi08tit} where it is shown
that for Poisson point processes, all connectivity properties are a function of
$\theta'=\theta^{d/\alpha}$ and do no depend on $\theta$ in any other way.

\subsection{$(1,1,1$): Partially fading random networks with slotted ALOHA}
If only the desired link is subject to fading (1/0 fading) and $\alpha=4$, we can
exploit \eqref{sousa_gen}, replacing $j\omega$ by $-\theta$,
to get
\begin{equation}
p_s^{1/0}=\L_I(\theta)=e^{-p\pi\Gamma(1-2/\alpha)\theta^{2/\alpha}}\,.
\label{sousa_ps_g}
\end{equation}
For $\alpha=4$,
\begin{equation}
p_s^{1/0}=\L_I(\theta)=e^{-p\sqrt{\theta}\pi^{3/2}}\,.
\label{sousa_ps}
\end{equation}
So $\gamma=\pi\Gamma(1-2/\alpha)\theta^{2/\alpha}$ which is larger than for the case with no fading
at all. So, as in the single-interferer case, it hurts the desired link if interferers do not fade.

\subsection{$(1,0,1)$: Non-fading random networks with $\alpha=4$ and slotted ALOHA}
From \cite[Eqn. (21)]{net:Sousa90}, $I^{-1}$ has the cdf
\begin{equation}
F_{I^{-1}}(\theta)=
\P[1/I < \theta]=1-p_s=\erf\left(\frac{\pi^{3/2}p\sqrt{\theta}}{2}\right)\,,
\label{outage_non}
\end{equation}
which is the outage probability for non-fading channels for a
transmitter-receiver distance 1.
For the spatial contention we obtain $\gamma=\pi\sqrt{\theta}$, and
it can be verified (\eg, by comparing Taylor expansions) that
$1-\gamma p < p_s(p) < \exp(-\gamma p)$ holds.

\subsection{$(1,1,1)$: Fully random networks with exponential path loss}
In \cite{net:Franceschetti04tap} the authors made a case for exponential path loss laws.
To determine their effect on the spatial contention,
consider the exponential path loss law $\exp(-\delta r)$ instead of $r^{-\alpha}$.
Following the derivation in \cite{net:Baccelli06}, we find
\begin{align}
  p_s&=\exp\left(-2\pi p\int_0^\infty \frac{r}{1+\exp(\delta r)/\theta}\d r\right)\nonumber\\
  &=\exp\left(-2\pi p\frac{-\dilog(\theta+1)}{\delta^2}\right)\,,
\end{align}
where $\dilog$ is the dilogarithm function defined as $\dilog(x) =  \int_1^x \log t/(1-t)\d t$.
So $\gamma=-2\pi\dilog(\theta+1)/\delta^2$.
The (negative) $\dilog$ function is bounded by
$-\dilog(x)<\log(x)^2/2+\pi^2/6$ \cite{net:Hassani07}, so 
\begin{equation}
 \gamma < \frac{\pi}{\delta^2}\left( \log^2(1+\theta)+\frac{\pi^2}{3}\right)\,,
\end{equation}
indicating that the spatial contention grows more slowly (with $\log\theta$ instead of
$\theta^{2/\alpha}$)
for large $\theta$ than for the power path loss law.
In the exponential case, finiteness of the integral is guaranteed for any $\delta>0$, in contrast 
to the power law where $\alpha$ needs to exceed the number of network dimensions.
Practical path loss laws may include both an exponential and a power law part, \eg,
$r^{-2}\exp(-\delta r)$. There are, however, no closed-form solutions for such path loss laws,
and one has to resort to numerical studies.

\section{Networks with Deterministic Node Placement}
In this section, we assume that $n$ interferers are placed
at fixed distances $r_i$ from the intended receiver.

\subsection{$(0,1,1)$: Fading networks with slotted ALOHA}
In this case, $p_s=\P[S \geqslant \theta I]$ for $I=\sum_{i=1}^n S_i
r_i^{-\alpha}$ and $S_i$ iid exponential with mean 1. 
For general $r_i$ and $\alpha$, we obtain from $p_s=\E[e^{-\theta I}]=\L_I(\theta)$ 
\begin{equation}
 p_s
= \prod_{i=1}^n \Bigl(1-\frac{p}{1+\xi_i}\Bigr)
\label{general_ps}
\end{equation}
where $\xi_i=r_i^\alpha/\theta$ is the effective distance. We find for the spatial
contention
\begin{equation}
\gamma\triangleq -\frac{\d p_s(p)}{\d p}\Big|_{p=0}=\sum_{i=1}^n \frac{1}{1+\xi_i}\,.
\label{gamma}
\end{equation}
Since $\d p_s/\d p$ is decreasing,
$p_s(p)$ is convex, so $1-p\gamma$ is a lower bound on the success probability.
On the other hand, $e^{-p\gamma}$ is an upper bound, since
\begin{equation}
 \log p_s=\sum_{i=1}^n \log\left(1-\frac{p}{1+\xi_i}\right)\lessapprox 
   \sum_{i=1}^n -\frac{p}{1+\xi_i}\,.
   \label{exp_bound}
\end{equation}
The upper bound is tight for small $p$ or $\xi_i$ large for most $i$, \ie, if most
interferers are far.

\subsection{$(0,1,1)$: Infinite regular line networks with fading and ALOHA}
Here we specialize to the case of regular one-dimensional (line) networks, where
$r_i=i$, $i\in\mathbb N$.

For $\alpha=2$, we obtain from \eqref{mathar} (or by direct calculation of \eqref{gamma})
\begin{equation}
 \gamma=\frac12 \left(\pi\sqrt\theta \coth(\pi\sqrt\theta)-1\right)\,.
 \label{gamma2}
\end{equation}
Since $x\coth x-1<x<x\coth x$, this is bounded by $(\pi\sqrt\theta-1)/2<\gamma<\pi\sqrt\theta/2$,
with the lower bound being very tight as soon as $\theta>1$. Again the success probability
is bounded by $1-\gamma p<p_s(p)<\exp(-p\gamma)$, and both these
bounds become tight as $\theta\rightarrow 0$,
and the upper bound becomes tight also as $\theta\rightarrow\infty$.

For $\alpha=4$, we first establish the analogous result to \eqref{mathar}.

\begin{proposition}
For one-sided infinite regular line networks ($r_i=i$, $i\in\mathbb{N}$) with slotted ALOHA and $\alpha=4$,
\begin{equation}
  p_s=\frac{\cosh^2\left(y(1-p)^{1/4}\right)
-\cos^2\left(y(1-p)^{1/4}\right)}{\sqrt{1-p}\,
(\cosh^2 y-\cos^2 y)}\,
\label{beta4_exact}
\end{equation}
where $y\triangleq \pi\theta^{1/4}/\sqrt 2$.
\end{proposition}
\begin{IEEEproof}
Rewrite \eqref{general_ps} as
\begin{equation}
 p_s=\frac{\prod_{i=1}^n (1+(1-p)\theta/i^4)} {\prod_{i=1}^n
 (1+\theta/i^4)}\,.
\end{equation}
The factorization of both numerator and denominator according to
$(1-z^4/i^4)=(1-z^2/i^2)(1+z^2/i^2)$ permits the use of Euler's
product formula $\sin(\pi z)\equiv \pi z\prod_{i=1}^\infty
(1-z^2/i^2)$ with $z=\sqrt{\pm j}((1-p)\theta)^{1/4}$ (numerator)
and $z=\sqrt{\pm j}\theta^{1/4}$ (denominator).  The two resulting
expressions are complex conjugates, and 
$|\sin(\sqrt{j}x)|^2=\cosh^2(x/\sqrt{2})-\cos^2(x/\sqrt{2})$.
\end{IEEEproof}
The spatial contention is
\begin{equation}
\gamma=\frac18 \frac{(y-1)e^{2y}+4\cos^2 y+4y\cos y\sin y-2-(y+1)e^{-2y}}{\cosh^2y-\cos^2y}\,.
\label{gamma4_exact}
\end{equation}
For $y\gtrapprox 2$, the $e^{2y}$ (numerator) and $\cosh^2 y$ (denominator) terms dominate,
so $\gamma\approx (y-1)/2$ for $y>2$. In terms of $\theta$, this implies that
\begin{equation}
  \gamma\approx \pi\theta^{1/4}/(2\sqrt 2)-1/2\,,
  \label{gamma4_approx}
\end{equation}
which is quite accurate as soon as $\theta>1$.
The corresponding approximation
\begin{equation}
 p_s\approx e^{-p\left(\pi\theta^{1/4}/(2\sqrt 2)-1/2\right)}\,.
 \label{beta4}
\end{equation}
can be derived from \eqref{beta4_exact} noting that for $y$ not too small
and $p$ not too close to $1$, the $\cosh$ terms dominate the $\cos$ terms and
$\cosh^2(x)\approx e^{2x}/4$, $1-(1-p)^{1/4}\approx p/4$, and
$(1-p)^{-1/2}\approx e^{p/2}$.

For general $\alpha$,
the Taylor expansion of \eqref{gamma} yields
\begin{equation}
 \gamma(\theta)=-\sum_{i=1}^\infty (-1)^i \zeta(\alpha i)\theta^i\,.
\end{equation}
In particular, $\gamma<\zeta(\alpha)\theta$.
Since $\zeta(x)\gtrapprox 1$ for $x>3$, the series converges quickly
for $\theta< 1/2$. For $\theta>1$, it is unsuitable.

\subsection{$(0,1,1)$: Partially fading regular networks}
If only the desired link is subject to fading, the success probability
is given by
\begin{equation}
 p_s=e^{-p\theta\sum_{i=1}^n r_i^{-\alpha}}\,,
\label{ps_partial}
\end{equation}
thus $\gamma=\sum_{i=1}^n 1/\xi_i$. Compared with \eqref{gamma},
$1+\xi$ is replaced by $\xi$. So the spatial contention is larger than in the case of
full fading, \ie, fading in the interferer's channels helps, as in the single-interferer case.

For regular line networks $\xi_i=i^\alpha/\theta$, so $\gamma=\theta\zeta(\alpha)$ and
$p_s=e^{-p\theta\zeta(\alpha)}$.

\subsection{$(0,1,0)$: Regular line networks with fading and TDMA}
\begin{figure}
\centering
\psfrag{T}{T}
\psfrag{R}{R}
\epsfig{file=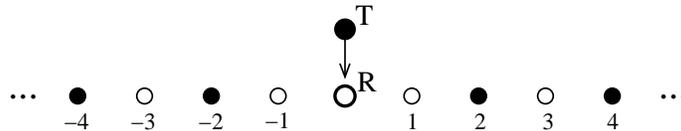,width=.55\columnwidth}
\caption{Two-sided regular line network with TDMA with $m=2$, \ie, every second node
transmits. The filled circles indicate the transmitters. The transmitter denoted by T is the
intended transmitter, the others are interferers. The receiver at the origin, denoted by R, is the
intended receiver. In the one-sided case, the nodes at positions $x<0$ do not exist.}
\label{fig:line_net}
\end{figure}

If in a TDMA scheme, only every $m$-th node transmits, the relative
distances of the interferers are increased by a factor of $m$. \figref{fig:line_net}
shows a two-sided regular line network with $m=2$. Since
$(mr)^\alpha/\theta=r^\alpha/(\theta m^{-\alpha})$, having
every $m$-th node transmit is equivalent to reducing the threshold
$\theta$ by a factor $m^\alpha$ and setting $p=1$.
\vspace*{-3mm}
\begin{proposition}
The success probability for one-sided infinite regular line networks with Rayleigh fading and $m$-phase TDMA is:
For $\alpha=2$:
\begin{equation}
  p_s=\frac{y}{\sinh y}\,,
\quad\text{\em where }y\triangleq\frac{\pi\sqrt{\theta}}{m}\,,
\label{unilat2}
\end{equation}
and for $\alpha=4$:
\begin{equation}
 p_s=\frac{2 y^2}{\cosh^2 y - \cos^2 y}\,,\quad\text{\em where }y\triangleq
\frac{\pi\theta^{1/4}}{\sqrt{2}m}\,.
\label{unilat4}
\end{equation}
\end{proposition}
\begin{IEEEproof}
Apply L'H\^opital's rule for $p=1$ in \eqref{mathar} and \eqref{beta4_exact}
(for $\alpha=2,4$, respectively) and replace
$\theta$ by $\theta m^{-\alpha}$.
\end{IEEEproof}

The following proposition establishes sharp bounds for arbitrary $\alpha$.
\vspace*{-3mm}
\begin{proposition}
The success probability
for one-sided infinite regular line networks, Rayleigh fading, and
$m$-phase TDMA is bounded by
\begin{equation}
   e^{-\zeta(\alpha)\theta/m^\alpha}\lessapprox  p_s \lessapprox\frac{1}{1+\zeta(\alpha)\frac{\theta}{m^\alpha}}\,.
\label{propo_upper}
\end{equation}
A tighter upper bound is
\begin{equation}
   p_s\lessapprox \frac{1}{1+\zeta(\alpha)\frac{\theta}{m^\alpha}+(\zeta(\alpha)-1)\frac{\theta^2}{m^{2\alpha}}} \,.
\label{propo_tight}
\end{equation}
\end{proposition}
\begin{IEEEproof}
Upper bound: We only need to proof the tighter bound.
Let $\theta'\triangleq\theta/m^\alpha$. The expansion of the product \eqref{general_ps},
$p_s^{-1}=\prod_{i=1}^\infty 1+\theta'/i^\alpha$\,,
ordered according to powers of $\theta'$, has only positive terms and starts with
$1+\theta'\zeta(\alpha)+\theta'^2(\zeta(\alpha)-1)$. There are more terms with
$\theta'^2$, but their coefficients are relatively small, so the bound is tight.
The lower bound is a special case of \eqref{exp_bound}.
\end{IEEEproof}
Note that all bounds approach the exact $p_s$ as $\theta/m^\alpha$ decreases. 
Interestingly, for $\alpha=2,4$,
the upper bound \eqref{propo_upper} corresponds exactly to the expressions
obtained when the denominators in \eqref{unilat2} and \eqref{unilat4}
are replaced by their Taylor expansions of order $2\alpha$.
Higher-order Taylor expansions, however, deviate from the
tighter bound \eqref{propo_tight}.

The success probabilities $p_s'$ for two-sided regular networks
are obtained simply by squaring the probabilities for the one-sided networks, \ie,
$p_s'=p_s^2$. This follows from the fact that the distances are related as follows:
$r_i'=r_{\lceil i/2\rceil}$.

\subsection{Spatial contention in TDMA networks}
In order to use the spatial contention framework for TDMA networks,
Let $\tilde p\triangleq 1/m$ be the fraction of time a node transmits.
Now $\d p_s/\d\tilde p|_{\tilde p=0}=0$ since $p_s$ depends on $m^\alpha$ rather
than $m$ itself. So for TDMA, we define
\begin{equation}
  \gamma\triangleq-\frac{\d p_s}{\d (\tilde p^\alpha)}\Big|_{\tilde p=0}
\end{equation}
and find $\gamma=\zeta(\alpha)\theta$, which is identical to the spatial
contention of the ALOHA line network with non-fading interferers.

Table \ref{table:gamma} summarizes the results on the spatial contention established
in this section.

\begin{table*}[ht]
\normalsize
\setlength{\extrarowheight}{2pt}
\[
\begin{array}{|c|c|c|c|l|}
\hline
\text{Uncertainty} & \text{Spatial contention }  \gamma & \text{Eqn.} & \text{\#dim.} & \text{Remark}\\[4pt]\hline
(1,1,1) & 2\pi\theta^{1/\alpha}\csc(\pi/\alpha)/\alpha & \eqref{ps-oned} & 1 & \text{Two-sided network}\\
& 2\pi^2\theta^{2/\alpha}\csc(2\pi/\alpha)/\alpha & \eqref{baccelli} & 2 & \text{From \cite{net:Baccelli06}.}\\
 & \pi^2\sqrt{\theta}/2 & \eqref{baccelli} & 2 & \text{Special case for }\alpha=4\\
 & \pi\Gamma(1-2/\alpha)\theta^{2/\alpha} & \eqref{sousa_ps_g} & 2 & \text{Non-fading interferers}\\
 & \pi^{3/2}\sqrt{\theta} & \eqref{sousa_ps} & 2 & \text{For }\alpha=4 \text{ and non-fading interferers}\\[2pt]\hline
(1,0,1) & \pi\sqrt{\theta} & \eqref{outage_non} & 2 & \text{No fading, for }\alpha=4 \\[2pt]\hline
(0,1,1) & \sum_{i=1}^{n} 1/(1+\xi_i) & \eqref{gamma} & d &\text{Deterministic node placement, $n$ nodes}\\
  & \pi\sqrt{\theta}\coth(\pi\sqrt{\theta})/2-1/2 & \eqref{gamma2} & 1 & \text{One-sided regular network, }\alpha=2\\
 & \approx\pi\theta^{1/4}/(2\sqrt{2})-1/2 & \eqref{gamma4_approx} & 1 & \text{One-sided regular network, }\alpha=4\\
 & \sum_{i=1}^{n} 1/\xi_i & \eqref{ps_partial} & d &\text{Det.~node placement, non-fading interf.}\\
 & \theta\zeta(\alpha) & \eqref{ps_partial} & 1 &\text{Regular network, non-fading interferers}\\[2pt]\hline
 (0,1,0) & p_s\gtrapprox e^{-\zeta(\alpha)\theta/m^\alpha}& \eqref{propo_upper} & 1 & \text{TDMA in one-sided regular networks.}\\[2pt]\hline
\end{array}
\]
\caption{Spatial contention parameters for different types of slotted ALOHA
networks.  For comparison, the TDMA case is added. ``Regular network" refers to an infinite line
network with unit node spacing.}
\label{table:gamma}
\end{table*}

\section{Throughput and Capacity}

\subsection{$(u_l,u_f,1)$: Networks with slotted ALOHA}
For networks with slotted ALOHA, define the {\em probabilistic throughput} as
\begin{equation}
\text{full-duplex:}\quad p_T^f\triangleq p\,p_s(p)\:;\qquad \text{half-duplex:}\quad p_T^h\triangleq p (1-p) \,p_s(p)\,.
\end{equation}
This is the unconditional probability of success, taking into account the probabilities
that the desired transmitters actually transmits and, in the half-duplex case, the desired
receiver actually listens.

\begin{proposition}[Maximum probabilistic throughput in ALOHA networks with fading]
Consider a network with ALOHA and Rayleigh fading with spatial contention $\gamma$ such that
$p_s=e^{-p\gamma}$. Then in the full-duplex case
\begin{equation}
 p_\opt=1/\gamma\,;\qquad p_{T\max}^f = \frac1{e\gamma}
\end{equation}
and in the half-duplex case
\begin{equation}
 p_{\opt}= \frac1\gamma+\frac{1}{2}\left(1-\sqrt{1+\frac4{\gamma^{2}}}\right)\,.
\label{popt}
\end{equation}
and
\begin{equation}
 p_{T\max}^h \gtrapprox \frac{1+\gamma}{(2+\gamma)^2}\exp\left(-\frac\gamma{2+\gamma}\right)\,,
 \label{pth_bound}
\end{equation}
\end{proposition}
\begin{IEEEproof}
Full-duplex: $p_\opt=1/\gamma$ maximizes $p\exp(-p\gamma)$.
Half-duplex:
Maximizing $\log p_T^h(p)$
yields the quadratic equation $p_\opt^2-p_{\opt}(1+2\sigma)+\sigma=0$ whose
solution is \eqref{popt}.
Any approximation of $p_\opt$ yields a lower bound on $p_T^h$. Since 
$p_\opt(0)=1/2$, and $p_\opt=\Theta(\gamma^{-1})$ for $\gamma\rightarrow\infty$,
a simple yet accurate choice is $p_\opt \gtrapprox 1/(2+\gamma)$ which results in the bound
in the proposition.
\end{IEEEproof}
Numerical calculations show that the lower 
bound \eqref{pth_bound} is within $1.4\%$ of the true maximum
over the whole range $\gamma\in\R^+$.

\subsection{$(0,1,0)$: Two-sided regular line networks with TDMA}
\begin{figure}
\centering
\epsfig{file=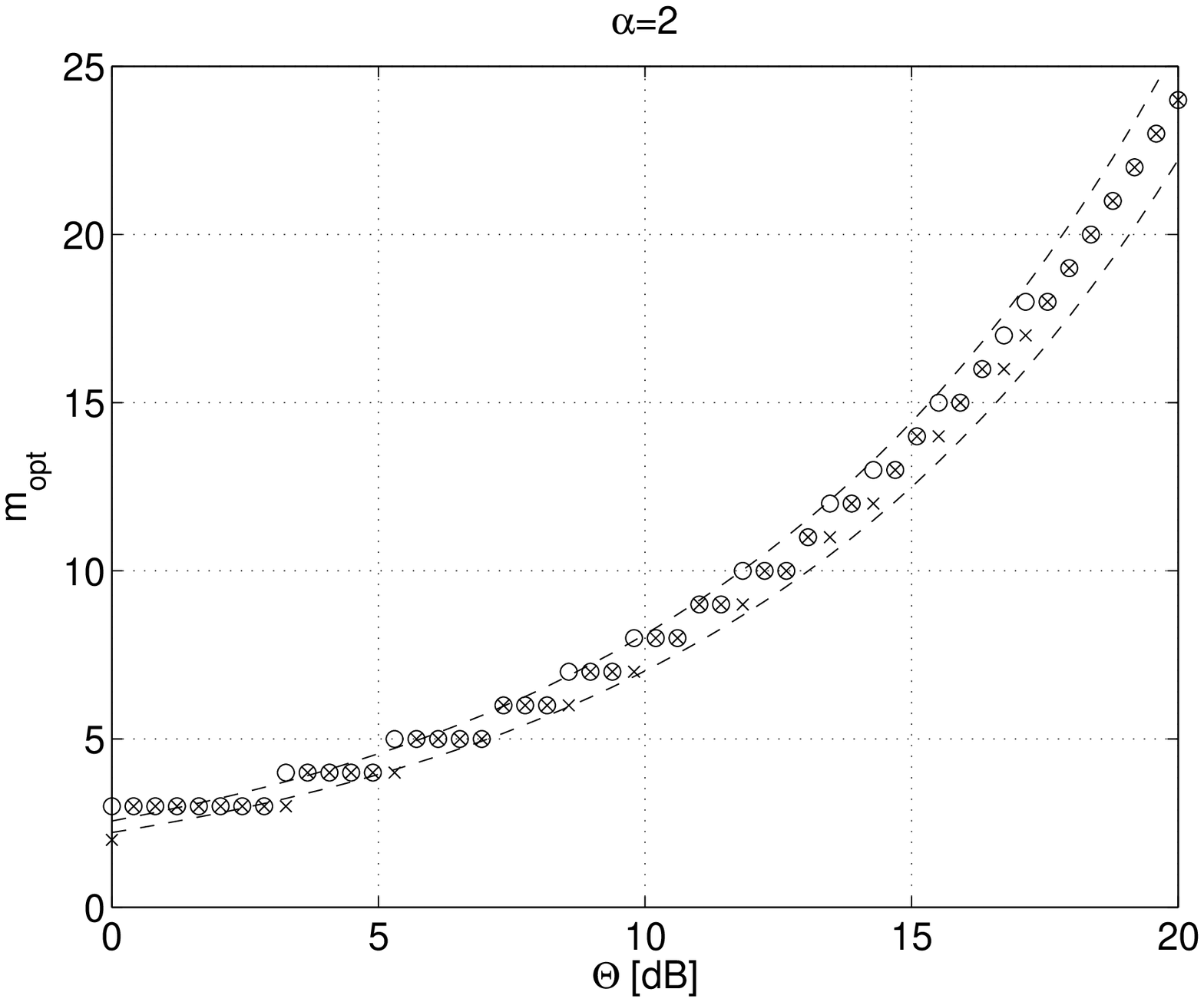,width=.48\columnwidth}\hfill
\epsfig{file=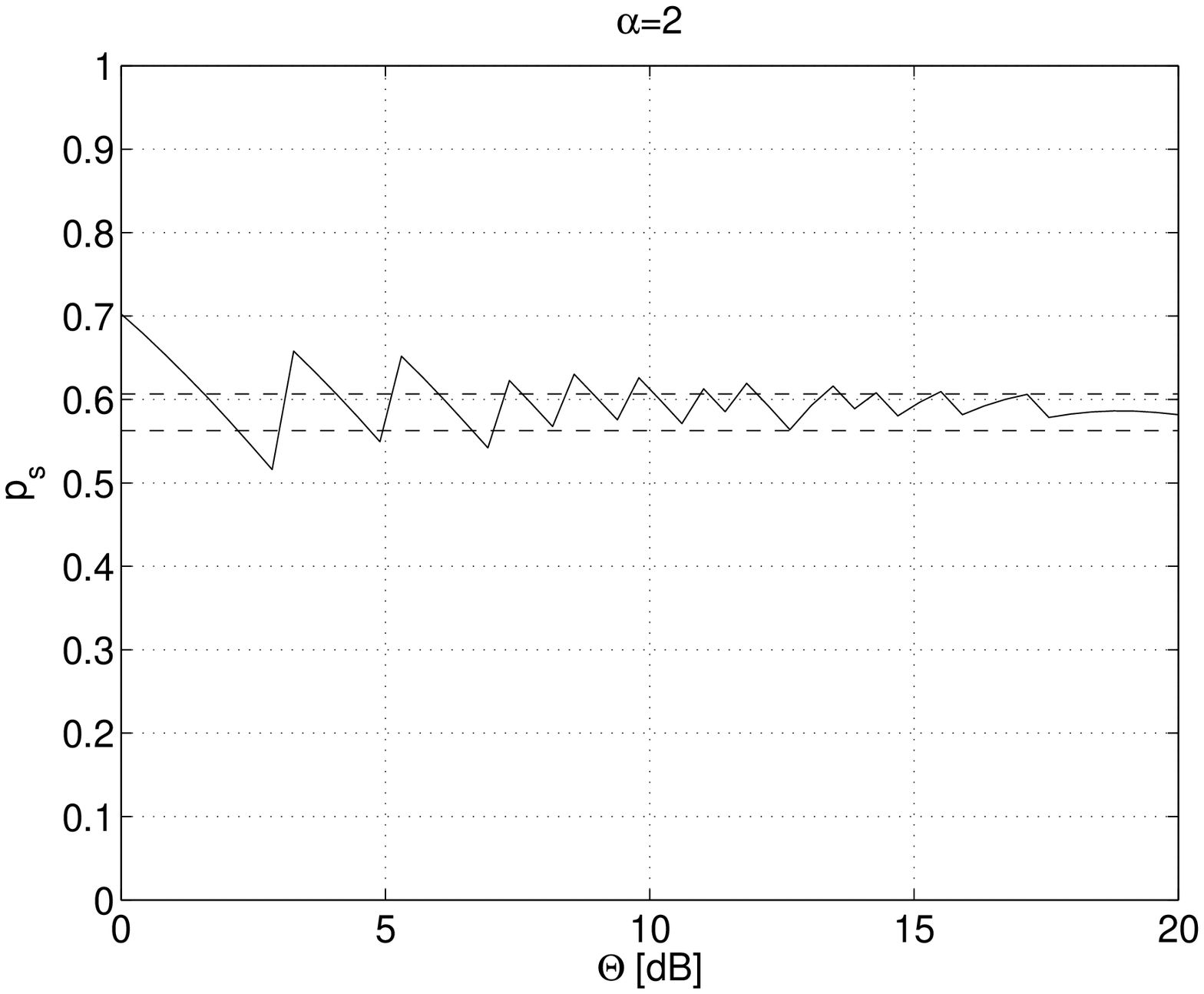,width=.48\columnwidth}
\caption{Left: Optimum TDMA parameter $m$ as a function of $\theta$ [dB] for 
$\alpha=2$. The dashed lines show the bounds \eqref{m_bounds}, the circles
indicate the true optimum $m_\opt$, the crosses the estimate $\hat{m}_\opt$ in \eqref{m_est}.
Right: $p_s$ for the optimum $m$ as a function of $\theta$ [dB] for $\alpha=2$. The dashed lines show the approximations \eqref{ps_approx}, the solid
line the actual value obtained numerically.}
\label{fig:m_values}
\end{figure}

Here we consider a {\em two-sided} infinite regular line network with
$m$-phase TDMA (see \figref{fig:line_net}).
To maximize the throughput $p_T\triangleq p_s/m$, we use the bounds
\eqref{propo_upper} for $p_s$. Since the
network is now two-sided, the expressions need to be
squared%
. Let $\tilde{m}_\opt\in\mathbb{R}$ and $\hat{m}_\opt\in\mathbb{N}$
be estimates for the true $m_\opt\in\mathbb{N}$.
We find
\begin{equation}
 \big(\theta\zeta(\alpha)(2\alpha-1)\big)^{1/\alpha} < \tilde{m}_\opt < \big(\theta\zeta(\alpha)2\alpha)\big)^{1/\alpha}\,,
\label{m_bounds}
\end{equation}
where the lower and upper bounds stem from maximizing the upper and lower
bounds in \eqref{propo_upper}, respectively. The factor 2 in $2\alpha$
indicates that the network is two-sided.
Rounding the average of the two bounds
to the nearest integer yields a good estimate for $m_\opt$:
\begin{equation}
\hat{m}_\opt=\lceil\big(\theta\zeta(\alpha)(2\alpha-1/2)\big)^{1/\alpha}\rfloor
\label{m_est}
\end{equation}
\figref{fig:m_values} (left) shows the bounds \eqref{m_bounds}, $\hat{m}_\opt$, and
the true $m_\opt$ (found numerically) for $\alpha=2$ as a function of
$\theta$. For most values of $\theta$, $\hat{m}_\opt=m_\opt$.
The resulting difference in the maximum achievable throughput $p_{T\max}$
is negligibly small. We can
obtain estimates on the success probability $p_s$ 
by inserting \eqref{m_bounds} into 
\eqref{propo_upper}:
\begin{equation}
  \left(1-\frac{1}{2\alpha}\right)^2 \approx p_s \approx  e^{-1/\alpha}\,.
\label{ps_approx}
\end{equation}
In \figref{fig:m_values} (right),
the actual $p_s(\theta)$ is shown with the two approximations for
$\alpha=2$. Since $m_\opt$ is increasing with $\theta$, the relative error
$\tilde{m}_\opt/m_\opt\rightarrow 0$, so we expect $\lim_{\theta\rightarrow\infty}p_s(\theta)$ to
lie between the approximations \eqref{ps_approx}.

\subsection{Rate optimization}
So far we have assumed that the SIR threshold $\theta$ is fixed and given.
Here we address the problem of finding the optimum rate of transmission for networks
where $\gamma\propto \theta^{d/\alpha}$, where $d=1,2$ indicates the number of
network dimensions.
We define the {\em throughput} as the product of the probabilistic throughput $p_T$
and the (normalized) rate of transmission $\log(1+\theta)$ (in nats/s/Hz).
As before, we distinguish the cases of half-duplex and full-duplex operation, \ie,
we maximize $p_T^f(\theta)\log(1+\theta)$ (full-duplex) or $p_T^h(\theta)\log(1+\theta)$
(half-duplex), respectively.

\begin{proposition}[Optimum SIR threshold for full-duplex operation]~\\
The throughput  $T=p\exp(-p\gamma)\log(1+\theta)$ is maximized at the
SIR threshold
\begin{equation}
\theta_\opt=\exp\bigg(\W\left(-\frac\alpha d e^{-\alpha/d}\right)+\frac\alpha d\bigg) -1\,,
\label{fullduplex}
\end{equation}
where $\W$ is the principal branch of the Lambert W function and $d=1,2$ is the number
of network dimensions.
\end{proposition}

\begin{IEEEproof}
 Given $\gamma$, the optimum $p$ is $1/\gamma$.
With $\gamma=c\theta^{d/\alpha}$, we need to maximize
\begin{equation}
  T(\alpha,\theta)=\frac1{e c \theta^{d/\alpha}} \log(1+\theta)\,,
\end{equation}
where $d=1,2$ is the number of dimensions.
Solving $\partial T/\partial\theta=0$ yields \eqref{fullduplex}.
\end{IEEEproof}
{\em Remark.}
$\theta_\opt$ in the two-dimensional case for a path loss exponent $\alpha$ equals
$\theta_\opt$ in the one-dimensional case for a path loss exponent $\alpha/2$.
In the two-dimensional case, the optimum threshold is smaller than one for
$\alpha<4\log 2\approx 2.77$.
  
The optimum (normalized) transmission rate (in nats/s/Hz) is 
\begin{equation}
 R_\opt(\alpha)=\log(1+\theta_\opt)=\W\left(-\frac\alpha d e^{-\alpha/d}\right)+\frac\alpha d\,,
 \quad d=1,2\,.
\end{equation}
$R_\opt(\alpha)$ is concave for $\alpha>d$, and the derivative at $\alpha=d$ is $2$ for $d=1$
and $1$ for $d=2$. So we have $R_\opt(\alpha)<\alpha-2$ for $d=2$ and $R_\opt(\alpha)<2
(\alpha-1)$ for $d=1$.

\begin{figure}
\epsfig{file=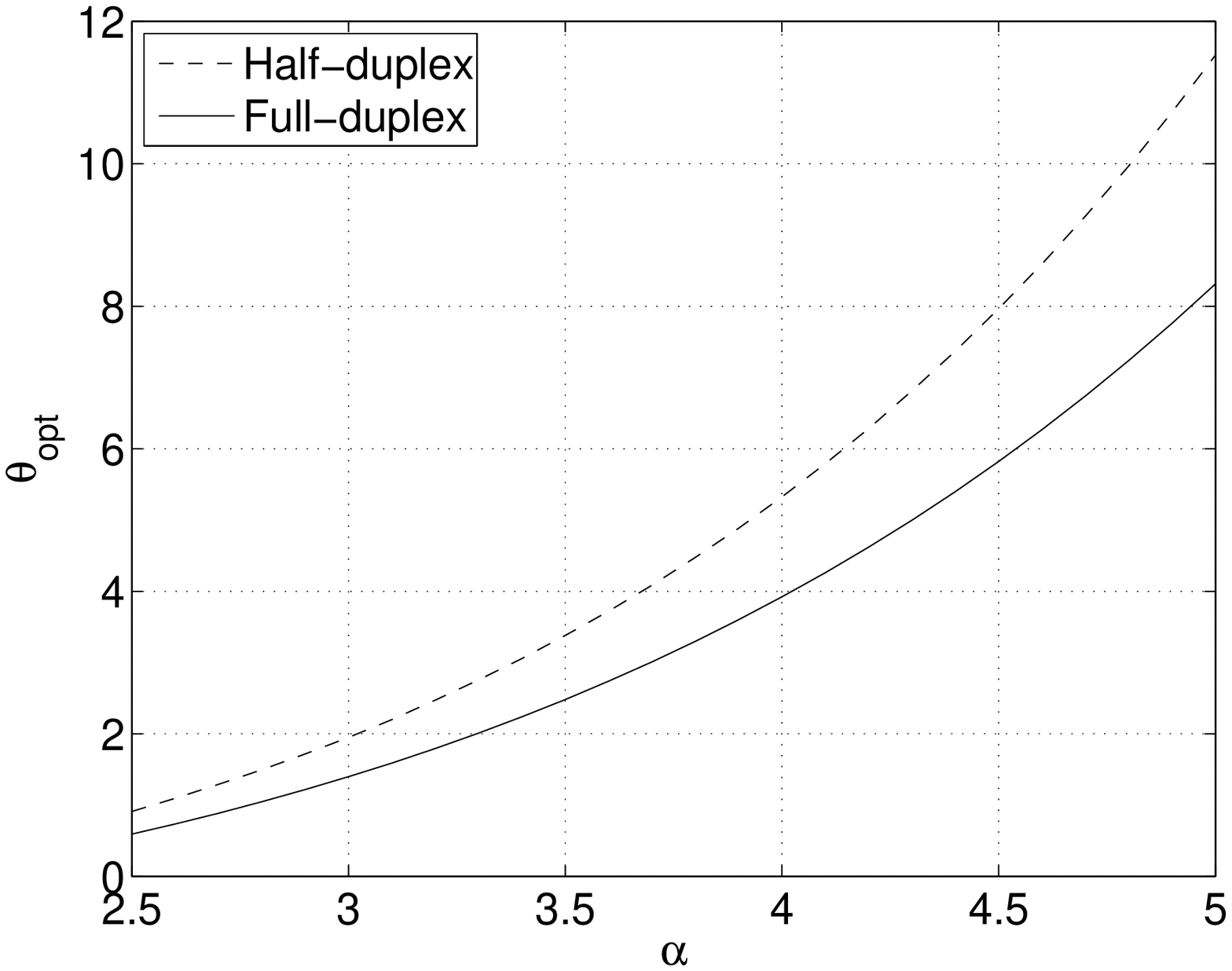,width=.48\columnwidth}\hfill
\epsfig{file=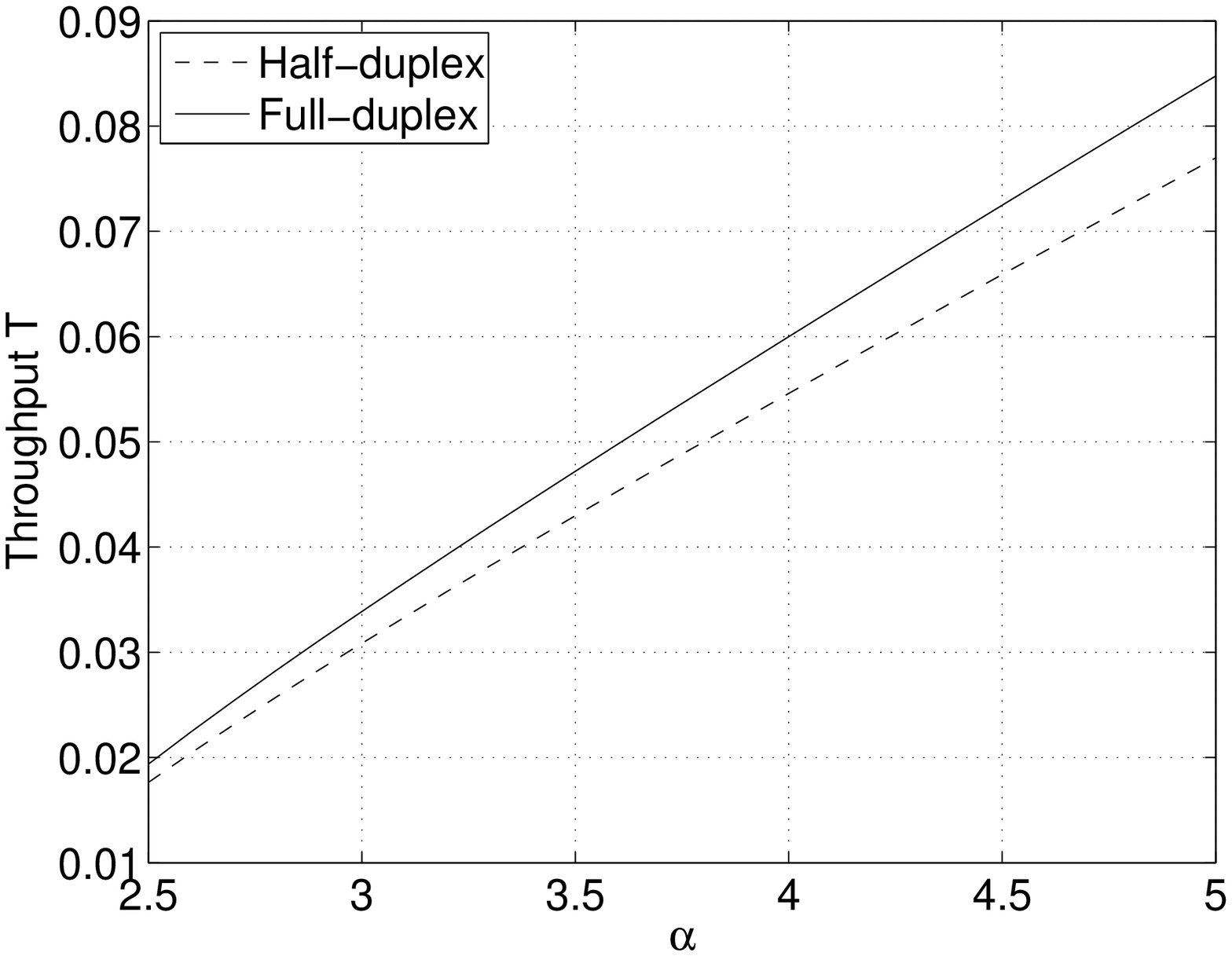,width=.48\columnwidth}
\caption{Left: Optimum threshold $\theta_\opt$ for full- and half-duplex operation as a function
of $\alpha$ for a two-dimensional network. Right: Maximum throughput.}
\label{fig:max_thru}
\end{figure}

In the half-duplex case, closed-form solutions are not available. The results of the numerical
throughput maximization are shown in \figref{fig:max_thru}, together with the results for the
full-duplex case. As can be seen, the maximum throughput scales almost linearly with
$\alpha-d$. The optimum transmit probabilities do not depend strongly on $\alpha$ and
are around $0.105$ for full-duplex operation and $0.08$ for half-duplex operation.
The achievable throughput for full-duplex operation is quite exactly 10\% higher than
for half-duplex operation, over the entire practical range of $\alpha$.

\subsection{$(1,1,1)$: Ergodic capacity}
Based on our definitions, the ergodic capacity can be generally
expressed as
\begin{equation}
  C=\E\log(1+\SIR)=\int_0^\infty -\log(1+\theta)\d p_s\,,
  \label{cap_general}
\end{equation}
where $p_s(\theta)$ is the ccdf of the SIR.

\begin{proposition}[Ergodic capacity for \boldmath$(1,1,1)$ networks]
Let $C$ be the ergodic capacity of a link in a two-dimensional $(1,1,1)$ network with 
transmit probability $p$.
For $\alpha=4$,
\begin{equation}
  C=2\Re\{q\}\cos(c_p)-2\Im\{q\}\sin(c_p)\,,
    \qquad q\triangleq\Ei(1,jc_p)\,,
    \label{erg_cap4}
\end{equation}
where $c_p=pC_2(\alpha)$ and $\Ei(1,z)=\int_1^\infty \exp(-xz)x^{-1}\d x$ is the exponential integral.
For general $\alpha>2$, $C$ is lower bounded as
\begin{multline}
 C > \log 2\cdot\Big(c_p^{-\alpha/2}\gamma(1+\alpha/2,c_p)+
   \left(\frac\alpha 4-1\right)\exp(-\sqrt 2c_p)+\exp(-c_p)\Big)+
    \frac{\alpha}{2}\Ei(\sqrt{2}c_p)\,,
    \label{erg_lower}
 \end{multline}
where  $\gamma(a,x)=\int_0^x t^{a-1}\exp(-t)\d t$ is the {\em lower} incomplete gamma function.

The one-dimensional network with path loss exponent $\alpha$ (and $c_p=pC_1(\alpha)$)
has the same capacity as the two-dimensional network with path loss exponent $2\alpha$.
\end{proposition}
\begin{IEEEproof}
Let $c_p\triangleq p\gamma\theta^{-2/\alpha}=pC_2(\alpha)$.
We have
\begin{align}
  C&=\frac{2c_p}{\alpha} \int_0^\infty \log(1+\theta)\theta^{2/\alpha-1}
    \exp(-c_p\theta^{2/\alpha})\d\theta\,\\
    &=c_p \int_0^\infty \log\left(1+t^{\alpha/2}\right) \exp(-c_pt)\d t\,.
    \label{erg_cap}
\end{align}
So, the $2/\alpha$-th moment of the SIR is exponentially distributed with mean $1/c_p$. As a consequence,
the capacity of the ALOHA channel is the capacity of a Rayleigh fading channel with mean SIR
$c_p^{-1}$ with an ``SIR boost'' exponent of $\alpha/2>1$. Note that since a significant part of
the probability mass may be located in the interval $0\leqslant\theta<1$, this does not mean
that the capacity is larger than for the standard Rayleigh case. This is only true if the SIR is
high on average.

For general $p$ and $\alpha$, the integral does not have a closed-form expression.
For $\alpha=4$, direct calculation of \eqref{erg_cap} yields
\begin{equation}
  C=\exp(-jc_p)\Ei(1,jc_p)+\exp(-jc_p)\Ei(1,-jc_p)\,,
\end{equation}
which equals \eqref{erg_cap4}.
To find an analytical lower bound, rewrite \eqref{erg_cap} as (by substituting $t\leftarrow t^{-1}$) 
\begin{equation}
 C= c_p\int_0^\infty \frac{\log(1+t^{-\alpha/2})\exp(-c_p/t)}{t^2}\d t
\end{equation}
and lower bound $\log(1+t^{-\alpha/2})$ by $L(t)$ given by
\begin{equation}
  L(t)=\begin{cases}
           -\frac\alpha2\log t & \text{ for }0\leqslant t < \sqrt{2}/2 \\
           \log 2 & \text{ for } \sqrt{2}/2\leqslant t < 1\\
           \log(2) t^{-\alpha/2} & \text{ for }1\leqslant t\,.
           \end{cases}
\end{equation}
This yields the lower bound \eqref{erg_lower}.
\end{IEEEproof}
For rational values of $\alpha$, pseudo-closed-form expressions are available using the
Meijer G function.

\begin{figure}
\centering
\epsfig{file=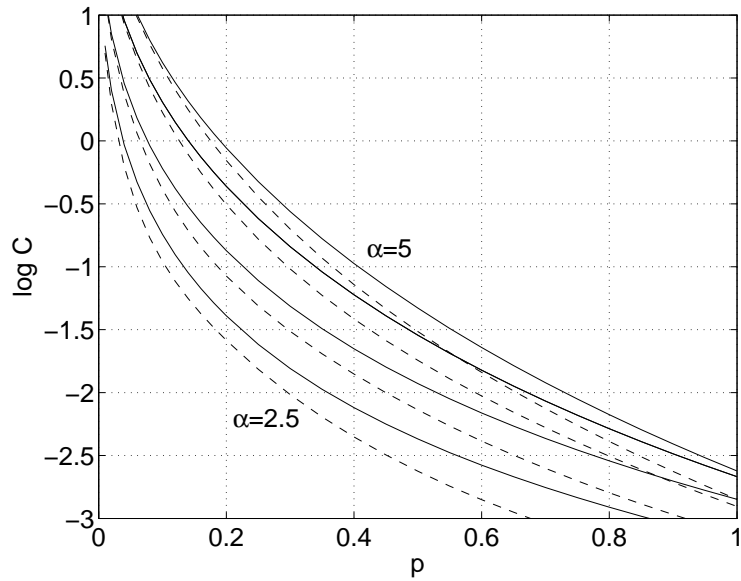,width=.6\columnwidth}
\caption{Ergodic capacity for a two-dimensional fading network with ALOHA for $\alpha=2.5,3,4,5$ as a function of $p$. The solid lines are the actual
capacities \eqref{erg_cap}, the dashed lines the lower bounds \eqref{erg_lower}.}
\label{fig:erg_capacity}
\end{figure}

\figref{fig:erg_capacity} displays the capacities and lower bounds for $\alpha=2.5,3,4,5$.
For small $c_p$ (high SIR on average), a simpler bound is
\begin{equation}
 C >  \int_1^\infty -\log(\theta)\d p_s= \frac\alpha 2\Ei(1,pC(\alpha))\,,
\end{equation}

\begin{figure}
\epsfig{file=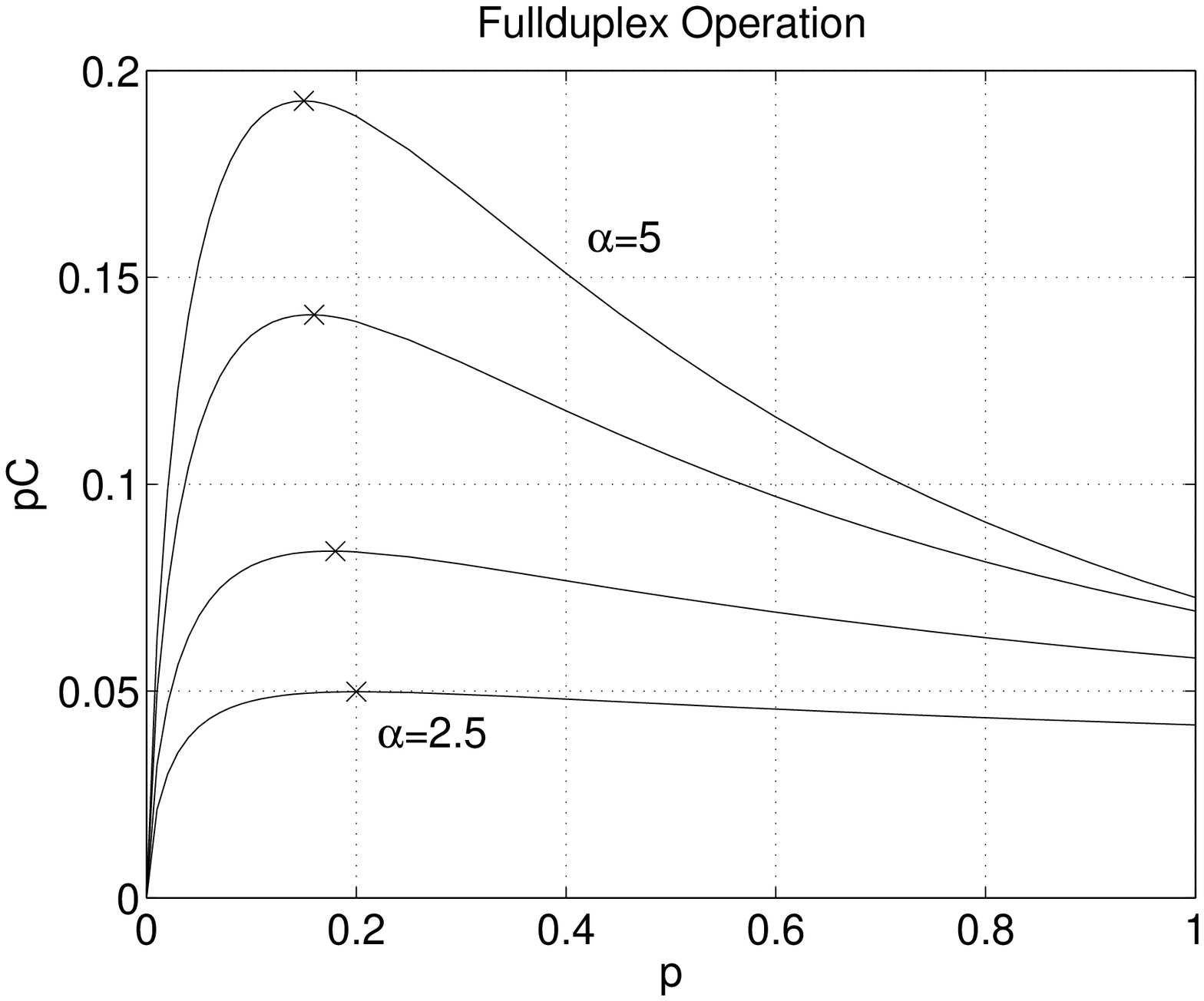,width=.48\columnwidth}\hfill
\epsfig{file=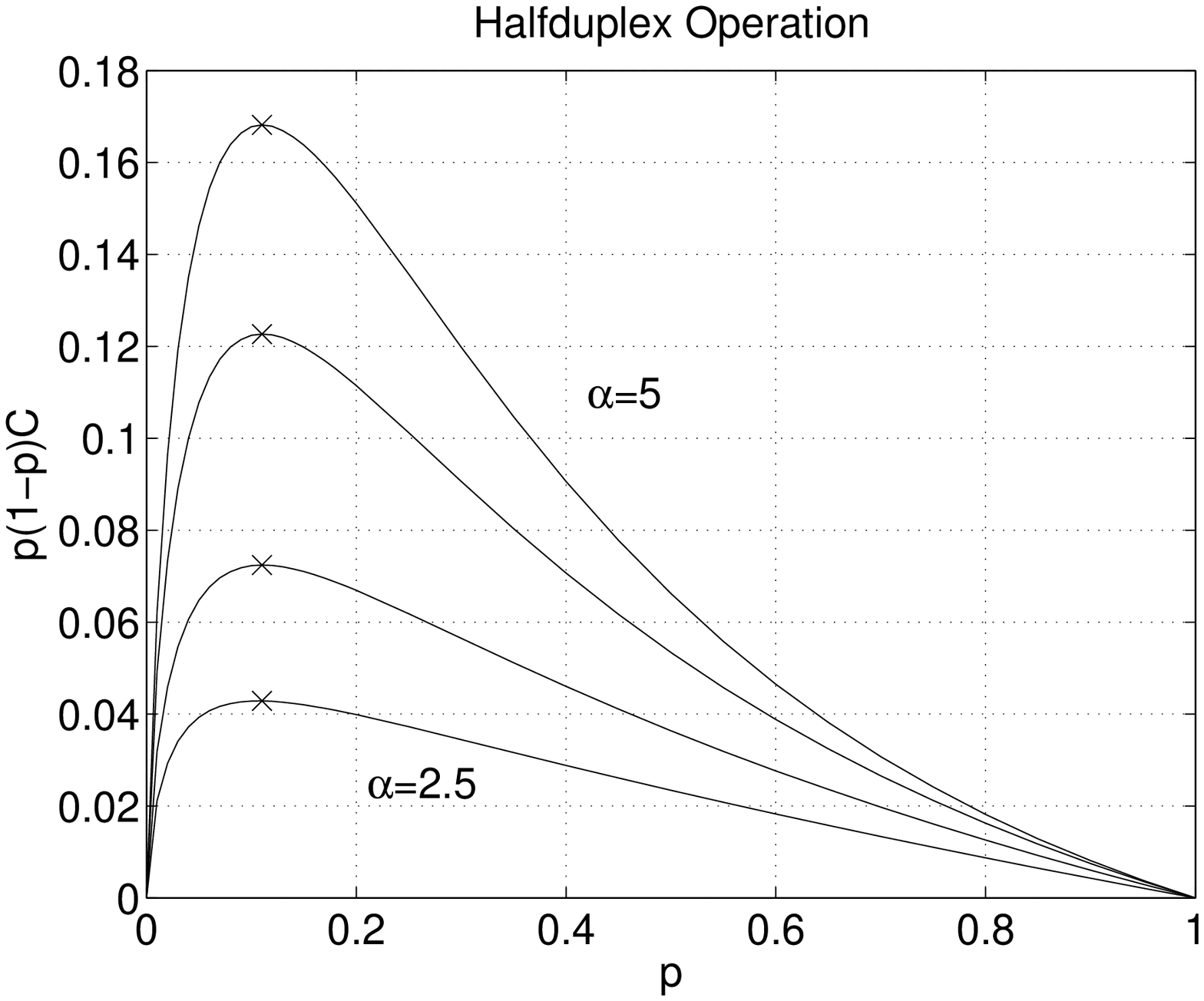,width=.48\columnwidth}
\caption{Spatial capacities for $\alpha=2.5,3,4,5$ as a function of $p$. Left plot: Full-duplex operation. Right plot: Half-duplex operation. The star marks the optimum $p$.}
\label{fig:duplex}
\end{figure}

To obtain the {\em spatial capacity}, the ergodic capacity needs to be multiplied
by the probability (density) of transmission. It is expected that there exists an optimum
$p$ maximizing the product $pC$ in the case of full-duplex operation or $p(1-p)C$ in the
case of half-duplex operation. The corresponding curves are shown in 
\figref{fig:duplex}. Interestingly, in the full-duplex case, the optimum $p$ is
{\em decreasing} with increasing $\alpha$. In the half-duplex case,
$p_\opt\approx 1/9$ quite exactly --- independent of $\alpha$.

\subsection{TDMA line networks}
\begin{figure}
\centering
\epsfig{file=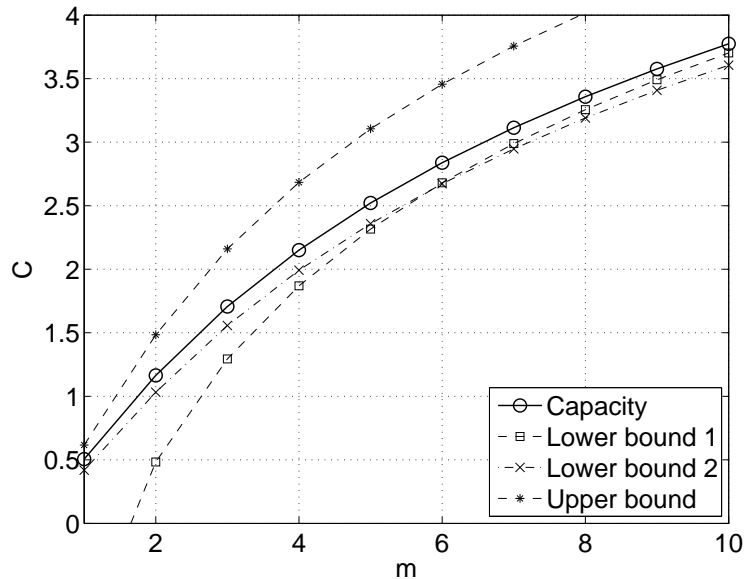,width=.6\columnwidth}
\caption{Ergodic capacity for TDMA line network for  $\alpha=2$ as a function of the reuse
parameter $m$. The solid line is the actual
capacity \eqref{erg_cap}, lower bound 1 and the upper bound are from \eqref{erg_tdma_low1}, and lower bound 2 is  \eqref{erg_tdma_low}.}
\label{fig:cap_tdma}
\end{figure}

\begin{proposition}[Ergodic capacity bounds for TDMA line networks]
For $\alpha=2$,
\begin{equation}
 2\log\left(\frac{2m}{\pi}\right) < C < \log\left(1+\frac{7\zeta(3)}{\pi^2}m^2\right)
 \label{erg_tdma_low1}
\end{equation}
and
\begin{equation}
  \E\sqrt\SIR=\frac{\pi}{4}m\,;\qquad \E\SIR=\frac{7\zeta(3)}{\pi^2}m^2\,.
\end{equation}  
For general $\alpha>1$,
\begin{equation}
   C > e^{\zeta(\alpha)/m^\alpha}\Ei(1,\zeta(\alpha)/m^\alpha) 
  \label{erg_tdma_low}
\end{equation}
and
\begin{equation}
 \E\SIR > \frac1{\zeta(\alpha)}m^\alpha\,.
 \label{SIR_low}
\end{equation}
\end{proposition}
\begin{IEEEproof}
$\alpha=2$:
Using \eqref{cap_general} and \eqref{unilat2} and substituting $t\leftarrow \pi\sqrt\theta/m$ yields
\begin{equation}
 C=\int_0^\infty \log\left(1+\left(\frac{mt}{\pi}\right)^2\right) \frac{t\cosh t-\sinh t}{\sinh^2 t}\d t
\end{equation}
Replacing $\log(1+x)$ by $\log x$ results in the lower bound which gets tighter as $m$ increases.
It also follows that $\pi\sqrt{\SIR}/m$ is distributed as
\begin{equation}
 \P(\pi\sqrt{\SIR}/m<t)= \frac{e^{2t}-2te^t-1}{e^{2t}-1}\,
\end{equation}
from which the moments of the $\SIR$ follow. The upper bound in \eqref{erg_tdma_low1} stems
from Jensen's inequality.
General $\alpha$: Use the lower bound  \eqref{propo_upper} on $p_s$ and calculate directly.
\end{IEEEproof}
\figref{fig:cap_tdma} shows the ergodic capacity for the TDMA line network for $\alpha=2$,
together with the lower bounds \eqref{erg_tdma_low1} and \eqref{erg_tdma_low} and the
upper bound from \eqref{erg_tdma_low1}. As can be
seen, the lower bound specific to $\alpha=2$ gets tighter for larger $m$. Using the lower bound
\eqref{SIR_low} on the SIR together with Jensen's inequality would result in a good
approximation $C\approx\log(1+m^\alpha/\zeta(\alpha))$.

From the slope of $C(m)$ it can be seen that the optimum spatial reuse factor $m=2$
maximizes the spatial capacity $C/m$ for $\alpha=2$. For $\alpha=4$, $m=3$ yields
a slightly higher $C/m$. This is in agreement with the observation made in
\figref{fig:duplex} (left) that in ALOHA $p_\opt$ slightly decreases as $\alpha$
increases.

\section{Discussion and Concluding Remarks}
We have introduced the uncertainty cube to classify wireless networks according
to their underlying stochastic processes. For large classes of networks, the outage
probability $\P(\SIR<\theta)$ of a unit-distance link is determined by the spatial
contention $\gamma$. Summarizing the outage results:
\begin{itemize}
\item For $(1,u_f,1)$ networks (PPP networks with ALOHA), $\gamma\propto \theta^{d/\alpha}$.
With Rayleigh fading, $p_s=\exp(-p \gamma)$, otherwise $p_s\leqslant \exp(-p \gamma)$.

\item For regular line networks with ALOHA (a class of $(0,1,1)$ networks),
  $\gamma\approx c\theta^{d/\alpha}-1/2$.
So, the regularity is reflected in the shift in $\gamma$ by $1/2$, \ie, $\gamma$ becomes affine
in $\theta^{d/\alpha}$ rather than linear.

\item Quite generally, with the exception of deterministic networks without fading interferers, $\gamma$
is a function of $\theta$ only through $\theta^{d/\alpha}$ (see Table \ref{table:gamma}).

\item For regular line networks with $m$-phase TDMA (a class of $(0,1,0)$ networks),
  $p_s\approx \exp(-\tilde p^\alpha \zeta(\alpha)\theta)$, where
$\tilde p=1/m$. So the increased efficiency of TDMA scheduling in line networks is
reflected in the exponent $\alpha$ of $\tilde p$.
\end{itemize}
The following interpretations of $\gamma=\sigma^{-1}$ demonstrate the fundamental nature
of this parameter:
\begin{itemize}
\item $\gamma$ determines how fast $p_s(p)$ decays as $p$ increases from 0:
$\partial p_s/\partial p|_{p=0}=-\gamma$.
\item For any ALOHA network with Rayleigh fading,
   there exists a unique parameter $\gamma$ such that
   $1-p\gamma\leqslant p_s\leqslant \exp(-p\gamma)$. This parameter is what we call the
   spatial contention. From all the networks studied, we conjecture that this is true for general
   ALOHA networks.
\item In a PPP network, the success probability equals the probability
that a disk of area $\gamma$ around the receiver is
free from concurrent transmitters. So an {\em equivalent disk model} could
be devised where the interference radius is $\sqrt{\gamma/\pi}$. For a transmission
over distance $R$, the disk radius would scale to $R\sqrt{\gamma/\pi}$.
\item In full-duplex operation, the probabilistic throughput is
$p_T^f=pe^{-p\gamma}$, and $p_\opt=\min\{\sigma,1\}$. So the spatial efficiency
equals the optimum transmit probability in ALOHA, and $p_T^f=\sigma/e$. The
throughput is proportional to $\sigma$.
\item The transmission capacity, introduced in \cite{net:Weber05}, is defined as the maximum spatial density of concurrent transmission allowed given an outage constraint $\epsilon$.
 In our framework, for small $\epsilon$, $p_s=1-p\gamma=1-\epsilon$, so $p=\epsilon\sigma$. So the
 transmission capacity is proportional to the spatial efficiency.
\item Even if the channel access protocol used is different from ALOHA, the spatial contention
offers a single-parameter characterization of the network's capabilities to use space.
\end{itemize}
Using the expressions for the success probabilities $p_s$, we have determined the optimum ALOHA
transmission probabilities $p$ and the optimum TDMA parameter $m$ that maximize the
probabilistic throughput.

Further, $p_s(\theta)$
enables determining both the optimum $\theta$ (rate of transmission) and the ergodic
capacity. For the cases where $\gamma\propto \theta^{d/\alpha}$,
$\SIR^{d/\alpha}$ is exponentially distributed.
The optimum rates and the throughput are roughly linear in $\alpha-d$, the spatial
capacity is about $2.5\times$ larger than the throughput,
and the penalty for half-duplex operation is 10-20\%.
The optimum transmit probability $p_\opt$ is around 1/9 for both optimum throughput
(\figref{fig:max_thru}, right) and maximum spatial capacity (\figref{fig:erg_capacity}, right). The
mean distance to the nearest interferer is $1/(2\sqrt{p_\opt})=3/2$, so for optimum performance
the nearest interferer is, on average, 50\% further away from the receiver than the desired
transmitter.
In line networks with $m$-phase TDMA, $\E\SIR$ grows with $m^\alpha$.

The results obtained can be generalized for (desired) link distances other than one
in a straightforward manner.
Many other extensions are possible, such as the inclusion of power control and
directional transmissions, as well as node distributions whose uncertainty lies
{\em inside} the uncertainty cube. 

\section*{Acknowledgment}
The support of the U.S.~National Science Foundation (grants
CNS 04-47869, DMS 505624, and CCF 728763) and the DARPA/IPTO IT-MANET program
through grant W911NF-07-1-0028 is gratefully acknowledged.

\bibliographystyle{IEEEtr}
\bibliography{header,comm,net}

\end{document}